\documentclass[12pt,a4paper]{article}
\usepackage{graphicx}
\usepackage{amssymb}
\usepackage{amsmath}
\usepackage{amsthm}
\usepackage{amsbsy}
\usepackage{cite}
\usepackage{verbatim} 
\usepackage{capt-of}
\usepackage{arydshln}
\usepackage[usenames]{color}
\usepackage{lineno}

\newcounter{CorrectThis}

\newcounter{MyPoints}

\newcommand{\nextPoint}[1]{\vskip 0.3cm \noindent \textit{#1}:}

\begin{document}

\pagestyle{empty}

\null
\vskip 2.0cm
\begin{center}
{\large {\bf 
On recrystallization nucleation in pure aluminum
}}
\\
\vskip 1.0cm
A. Morawiec\\
Polish Academy of Sciences, 
Institute of Metallurgy and Materials Science \\
Reymonta 25, 30--059 Krak{\'o}w, Poland \\
E-mail: nmmorawi@cyf-kr.edu.pl  \\ 
Tel.: ++48--122952854 \ \ \  Fax: ++48--122952804 
\end{center}

\vfill 


\noindent
{\bf Abstract}
\\
Static recrystallization is an important aspect of metal processing. 
The initial stage of recrystallization -- nucleation of new grains -- 
determines its later stages. 
The accepted mechanisms of recrystallization nucleation
are based on the assumption that embryos with orientations of the nuclei preexist
in the deformed matrix. 
However, this standard picture 
seems to be incomplete.
There are indications that, 
in some cases,   
the deformed matrix has no orientations observed in recrystallized material.
Therefore a mechanism of early stage recrystallization without 
preexisting embryos is considered.
Since the recrystallization growth shows strong orientation selection, and 
the recrystallization front is believed to migrate
through collective shuffling of atoms,
it is postulated that the shuffling mechanism 
responsible for oriented growth of a recrystallizing grain
extends to the very beginning of the grain's existence, i.e.,
that a new orientation can be created 
via rearrangement of atoms in a strained region.
The postulate explains the formation of new orientations, and 
it has the potential to significantly change the understanding 
of the early phase of recrystallization.

\vskip 0.2cm

\noindent
\hrulefill 




\newpage
\pagestyle{plain}

\section{Introduction} 

Recrystallization, i.e., changes of microstructures and crystallographic textures 
that occur in metals annealed after plastic deformation,
is an important element of metal processing.
Like the preceding stage of recovery, recrystallization is driven by stored energy of the deformation, 
but as opposed to recovery, it involves formation and migration of grain boundaries
\cite{Humphreys_2004a}. 
The start of the recrystallization process is usually relatively slow. 
The apparent incubation period referred to as recrystallization nucleation
is followed by a faster recrystallization growth.
In simplified terms, 
nucleation comprises the development of small strain-free crystallites (nuclei)
capable of growth into the deformed matrix.
The boundaries between the aforementioned stages of annealing are not sharp;
one may observe dynamic recovery occurring during deformation, 
nucleation during recovery, overlap of the nucleation and growth 
stages and overlap of recrystallization and the subsequent 
stage of grain growth driven by grain boundary curvature.

Insight into
the initial phase of recrystallization is important  
for understanding  its later stages, and it is key for meaningful 
modeling and simulations of the entire process.
On the other hand, 
the nucleation is difficult to investigate 
due to its random nature and the rarity of the nuclei.
A grain is recognized as a nucleus only when it is relatively large,
and its origins and former surroundings are not reconstructible
by the currently available experimental techniques.

There are indications that the accepted nucleation theories
do not fully explain early stages of recrystallization.
The theories assume that embryos in orientations of recrystallization nuclei 
preexist in the deformed material.
However, some observations contradict this assumption. 
Therefore an addition to these theories is considered.
The paper aims to draw attention to the fact that
besides the transformation of preexisting embryos 
into nuclei, the option that nuclei rise
in highly strained regions by orientation-changing 
collective rearrangements of atoms
cannot be excluded.
It is postulated that the mechanism 
responsible for recrystallization growth 
extends to the beginning of existence of new grains, 
and orientations not observed in the deformed state arise 
via 
rearrangements of atoms in strained regions.
The paper does not report any new experimental data.
It only presents a new interpretation of available information.

Sections two and three outline relevant facts about recrystallization nucleation 
and oriented growth. 
Section four describes the `nucleation problem', 
i.e., a gap in the current theory of recrystallization nucleation.
A new postulate, which is a step towards filling the gap, is discussed in section five. 
Before proceeding to the merits, basics 
concerning the topic under consideration 
are recalled in the next three paragraphs.

\nextPoint{Pure aluminum as model case} 
The recrystallization process 
is influenced by numerous factors. 
These are in particular  
the crystal structure, stacking fault energy,
dissolved impurities, particles of another phase,
deformation path 
and the initial state of the sample. 
Considering annealing in pure aluminum allows for 
avoiding complications caused by second phase particles 
and 
those related to annealing twinning. 
The text below applies to this case, unless explicitly stated otherwise.

\nextPoint{Deformation structure}
Sufficiently large deformation of a polycrystal leads to a cell structure; e.g., \cite{Gay_1954}. 
Original grains separated by old grain boundaries are
divided into cells or subgrains of low dislocation density.
Cells are surrounded by dense dislocation tangles, i.e., diffuse intragranular boundaries
known as cell walls or subboundaries.
With growing deformation, the cell size decreases to a limiting value 
(of about $2\mu\text{m}$ in the case of Al).
During annealing, 
the cell structure is further sharpened by polygonization 
-- a constituent of the recovery process, 
and most of the energy of deformation is stored in the cell walls.

\nextPoint{Recrystallization texture} 
If the deformation exceeds a certain minimum critical strain, 
annealing at sufficiently high temperature will cause recrystallization.
The latter leads to a microstructure with strain-free grains and 
affects the orientation distribution.
The recrystallization texture can be significantly 
different from the deformation texture.
Hence, the natural question about the origin of the 
recrystallization textures:
are they a result of nucleation in selected orientations 
(oriented nucleation \cite{Burgers_1931,Burgers_1953}),
or are they formed by fast growth of grains in certain orientations
(oriented growth \cite{Beck_1950b,Beck_1953})?

\section{Views on recrystallization nucleation}

Most of the currently accepted ideas concerning recrystallization nucleation 
were formulated by the nineteen seventies.
For detailed descriptions the
theories of the nucleation mechanism, the reader is referred to
reviews on recrystallization
\cite{Cotterill_1976,Doherty_1997,Doherty_1997b,Humphreys_2004a,Rios_2005,Raabe_2014} 
and recrystallization nucleation 
\cite{Doherty_1974,Hutchinson_1992,Humphreys_2004,Knipschildt_2022}.
The theories are very briefly summarized below.

\nextPoint{Rejection of conventional nucleation mechanism} 
In the classical picture of thermally-activated nucleation 
in 
precipitation reactions, 
the growth of a nucleus is governed by the interplay between 
the areal free energy of the matrix-particle interface and 
the difference of volumetric free energies of the matrix and the particle \cite{Burke_1952}. 
This description involves critical size of a nucleus: 
only being larger than the critical size guarantees growth.
In the case of recrystallization, it is argued that 
with typical values of the stored energy of deformation and
grain boundary energy, the critical sizes are 
prohibitively large \cite{Beck_1954,Bailey_1960}. 
Therefore, it is generally accepted that new grains are formed 
in a way other than the classical nucleation.
It is commonly believed that embryos which then become nuclei 
are present in the deformed matrix, 
and all growing grains have orientations inherited from the embryos; 
see, e.g., 
\cite{Beck_1954,Doherty_1974,Humphreys_2004a,Doherty_1997,Doherty_1997b,Rios_2005,Raabe_2014}.
Accordingly, the term 'recrystallization nucleation' does not refer
to the classical model of formation of new structures via thermal fluctuations but 
to origins of the embryos and their transformation into growing grains.

\nextPoint{Block theories} 
As early as 1942, Burgers made a distinction between types of preexisting embryos, 
or deformed matrix 'blocks', as they were then called,
and also between theories based on the assumption of such blocks
\cite{Burke_1952,Beck_1954}. 
A `low-energy' (`high-energy') block is a volume  
in which the strain is smaller (larger) than average. 
The low-energy blocks with low density of defects are ready to expand, whereas
the high-energy blocks with high dislocation density 
recover early and quickly acquire ability to grow.
In other words,  
with the low-energy block theory, a grain survives deformation strain-free, 
and with the high-energy block theory,  a grain becomes strain-free at the early stage of annealing.

\nextPoint{SIBM} Beck and Sperry 
observed that some subgrains adjacent to grain boundaries 
grow into grains on the other side of the boundary \cite{Beck_1950a}.
With sufficiently large difference in stored energy over sufficiently large area 
of the boundary, low-energy or early recovered subgrains bulge into neighboring grains 
\cite{Bailey_1962}. 
The mechanism is known as `strain induced boundary migration' or SIBM.
Clearly, the new defect-free grain has the same orientation
as the subgrain from which it originates.

\nextPoint{Ledge theory}
Related to SIBM is  
the `ledge theory' \cite{Davies_1961}.
Deformation leads to ledges (slip steps) on high-angle grain boundaries, and 
nucleation occurs when a boundary near a ledge straightens out 
to reduce its area; 
the volume swept by the boundary is a defect-free recrystallization nucleus.

\nextPoint{Coalescence} 
Electron microscopic observations of growth of subgrains
by disappearance of the subboundaries led to the 
idea that nuclei appear via 
coalescence of adjacent subgrains; 
see, e.g., \cite{Hu_1962} and references therein. 
It was believed that the merger of subgrains into the same orientation
occurs via small-angle rotations of the coalescing cells \cite{Li_1962}.\footnote{It is 
worth noting that a rotation of a grain involves rearrangement 
of atoms over its entire surface,
and it is nowadays known that noticeable
thermally-activated rotations are possible only for small nanograins. 
}

\nextPoint{Cahn-Cottrell mechanism} 
Beck \cite{Beck_1949} and Cahn \cite{Cahn_1949}
proposed that nuclei are results of polygonization.
Polygonization causes additional accumulation of dislocations
in cell walls. That increases cell misorientations and makes the subboundaries
capable of migration \cite{Cottrell_1953}.

\nextPoint{Discontinuous subgrain coarsening} 
The growth of subgrains within a grain by  
migration of subboundaries was confirmed by TEM observations \cite{Fujita_1961}.
Experiments on deformed single crystals led to suggesting that
recrystallization is initiated in areas with large orientation gradients
by migration of low-angle boundaries of some subgrains \cite{Ferry_1996,Huang_2000}.
This mechanism is referred to as `discontinuous subgrain coarsening'.

\nextPoint{Operating early-stage recrystallization mechanisms} 
Taking into account the variety of factors influencing the course of recrystallization, 
it seems that no single nucleation mechanism is applicable to all 
deformation levels or annealing temperatures. 
The rates of the operating mechanisms
are different in different cases. 
Knowing the shares and rates (i.e., kinetics) of active mechanisms
is the essential to understanding the early phase of recrystallization.

\section{Oriented growth}

In the framework of the current recrystallization theory, 
with a thermodynamically stable nucleus already present in the deformed matrix,
the key condition for growth is having a mobile 
boundary.
In recrystallization, the most mobile 
boundaries correspond to specific misorientations.

\nextPoint{Growth selection} 
It has long been observed that the misorientations 
between recrystallized grains and the deformed matrix 
in Al are not random;
e.g., \cite{Beck_1949}.
Experiments on 
cold-rolled (e.g., \cite{Beck_1950b,Kohara_1958b}) 
and strained wire-shaped  (e.g., \cite{Yoshida_1959})
single crystals
showed strong growth selection.
These experiments do not exclude oriented nucleation, but they
prove that 
the recrystallization textures result, at least in part, 
from the oriented growth.

\begin{figure}[t]
\begin{picture}(300,220)(0,0)
\put(0,2){\resizebox{7.2 cm}{!}{\includegraphics{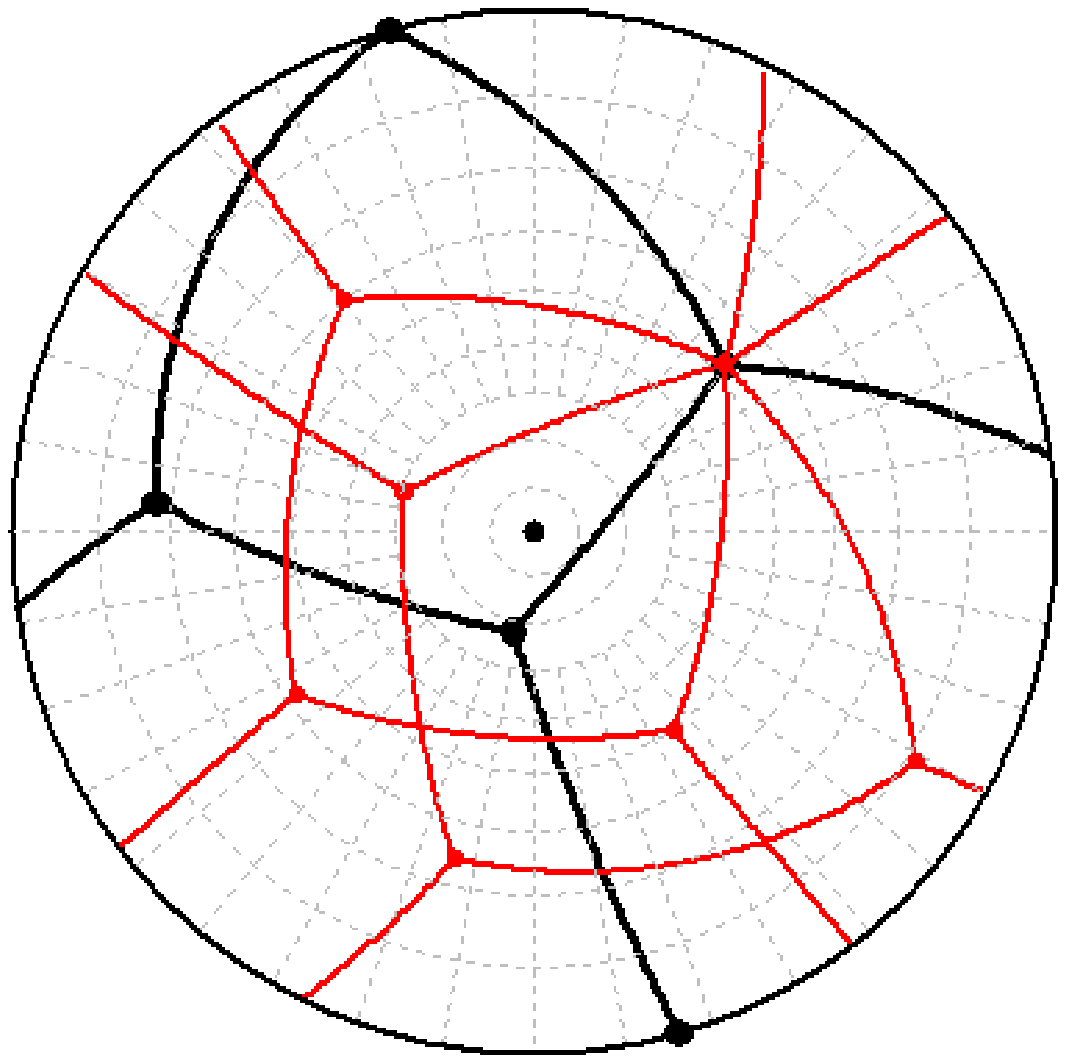}}}
\put(220,0){\resizebox{7.2 cm}{!}{\includegraphics{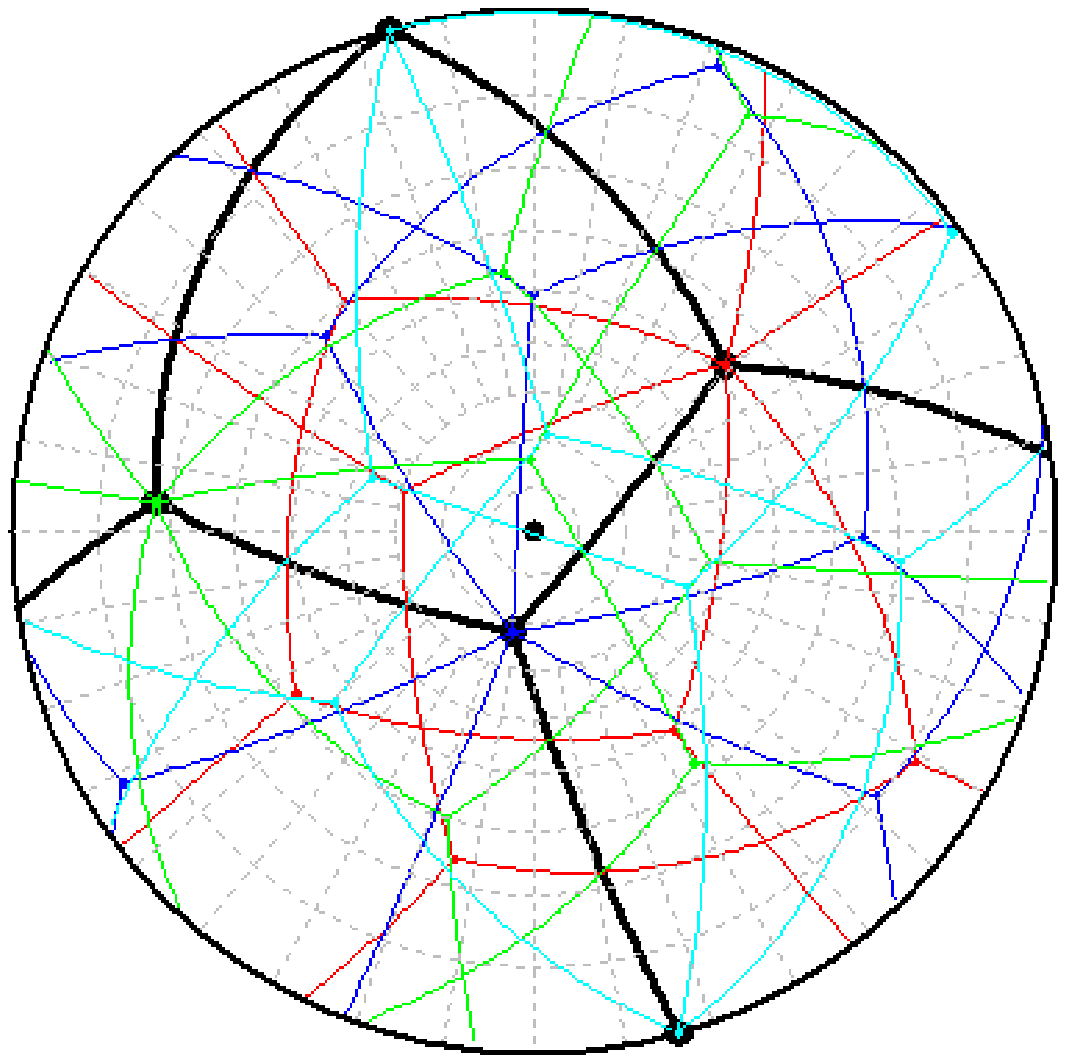}}}
\put(10,210){\textit{a}}
\put(230,210){\textit{b}}
\put(94,206){\small RD}
\put(314,206){\small RD}
\put(202,103){\small TD}
\put(423,103){\small TD}
\put(110,180){$S$}
\end{picture}
\vskip 0.0cm
\caption{Example illustrating orientations related by $M_{g}$ to 
the $(\overline{1}\,\overline{2}\,3)[634] \sim S$ orientation.
(\textit{a}) Stereographic projection of 111 pole figure with  
$S$ 
and two orientations related to it by $\pm 40^{\circ}$ rotations about one of the 
$\langle 111 \rangle$
axes. 
Note that one of the latter orientations is close to the cube orientation.
(\textit{b}) The pole figure with all eight orientation variants related to $S$
by $\pm 40^{\circ}$ rotations about all $\langle 111 \rangle$ axes.
In the case of rolled samples with the orthorhombic symmetry, 
there are four variants of $S$
and 
32 orientation variants resulting from the $M_{g}$ misorientation to $S$. 
}
\label{Fig_S_40_111}
\end{figure}

The misorientation of grains with the fastest recrystallization growth 
will be denoted by $M_{g}$.
Its axis is close to $\langle 111 \rangle$,
but estimations of the angle vary considerably.
Typically, this angle is assumed to be around $40^{\circ}$.
Due to crystal and sample symmetries, the single orientation relationship $M_{g}$
translates into a number of distinct orientations (variants) of the recrystallized matter; 
see Fig. \ref{Fig_S_40_111}.
The growth of grains with the $M_{g}$ misorientation  may 
be subject to variant selection; see Appendix.

\nextPoint{Oriented growth and tilt boundaries} 
Growth selection of the misorientation is a pronounced consequence of anisotropy. 
But it is not the only one.
Based on shapes of the fast growing grains with the near-$M_{g}$ misorientations, 
it was deduced that most mobile are the 
$\langle 111 \rangle$ tilt boundaries 
\cite{Lucke_1974,Gottstein_1978}.
With high mobility of planes parallel to a 
$\langle 111 \rangle$-type direction common for the grain and the matrix,
the grain attains a plate-like shape, 
and the plate is perpendicular to that direction. 
This geometry is experimentally observed; see, e.g., \cite{Lucke_1974}.

\nextPoint{General remark on boundary motion in pure metals} 
Grain boundary movement relies on transfer of atoms across the boundary.
Macroscopically, 
the rate of the transfer is governed by the driving force and boundary mobility. 
The vast majority of data on boundary mobility
concerns the curvature driven boundary migration, and 
some relate to the migration induced by elastic strain.
On the other hand, relevant for 
recrystallization is the boundary motion 
driven by stored energy of plastic deformation.

Various atomic-scale mechanisms of boundary migration have been considered, but they are
still not fully understood.
For a given material, the type of the prevailing mechanism may depend on temperature 
and the nature and magnitude of the driving force.
The basic way of migration is via single atom hopping 
(as in boundary self-diffusion).
A special boundary can be displaced by passages of grain boundary dislocations.
However, the mechanism which seems to be most relevant in the considered case 
is the collective shuffling (rearrangement) of atoms, i.e., 
a nearly instantaneous change of boundary position
caused by only small but cooperative displacement of 
a number of atoms.
Both experiments \cite{Babcock_1989b,Merkle_2002,Merkle_2004}
and simulations \cite{Jhan_1990,Schoenfelder_1997}
show that movements of some high-angle 
boundaries involve collective shuffles of clusters of atoms.
In some circumstances,
a cluster of several hundred atoms may  
switch its crystallographic orientation 
\cite{Merkle_2002}.
The shuffling mechanism is usually associated with 
faceted low-$\Sigma$ CSL boundaries, 
but clearly, it is not limited to these cases \cite{Bowers_2016}.
If a boundary moves due to a driving force,
the collective nature of atom rearrangements and 
short-range of atom displacements cause that 
the migration by shuffles is more anisotropic 
than the migration by 
hopping of individual atoms \cite{McCarthy_2020}.
Collective shuffles of large clusters of atoms
can make the boundary motion jerky.

\nextPoint{Mechanism of oriented growth}
Coming back to the recrystallization growth, 
the question is why the $40^{\circ}$ $\langle 111 \rangle$ tilt boundaries have 
mobility much higher than other boundaries. 
One on the past views 
was that the self-diffusion rate for
boundary edge dislocations is higher than that for
screw dislocations, and, in effect, 
the self-diffusion rate along tilt boundaries with a large share of edge
dislocations would be higher than that along twist boundaries with a large share of screw
dislocations  \cite{Kohara_1958}.

However, it seems that in the case of oriented recrystallization growth,
the single atom hopping is dominated by shuffling of clusters of atoms.
Shuffling occurs even in static boundaries between defect-free crystals \cite{Merkle_2002}, 
whereas the mobile recrystallization front separates a crystal and 
a strained matrix, so shuffles in the matrix are more likely. 
Moreover, the shuffling mechanism is clearly indicated by the anisotropy of mobility
(exhibited by orientation selection and the tilt character of mobile interfaces). 
Migration by collective movements of atoms is indirectly corroborated by 
jerky motion
of the recrystallization front 
observed in TEM.\footnote{In this case, 
however, one must take into account that these are thin-foil experiments 
in which boundary movements are affected by surface pinning.} 

Regardless of the detailed nature of the transfer of atoms under the stored energy driving force, 
the mechanism is anisotropic, which indicates that it is collective.
Further on, an elementary constituent of this mechanism will be referred to as 
`stored-energy (driven) collective rearrangement (of atoms)' which 
will be abbreviated to SECR.

\nextPoint{SECR versus conventional grain boundary shuffling} 
The rearrangement of atoms occurring during oriented recrystallization growth and 
involving consumption of strained regions is different than the 
primary shuffling of Merkle \cite{Merkle_2002,Merkle_2004}.
The latter transforms a 'part of a crystal' to a 'part of a crystal', 
whereas the collective rearrangement induced by strain transforms a 'part of a 
strained crystal' to a 'part of a strain-free crystal'. 
In the absence of a driving force, the Merkle's shuffling 
does not involve an energy change and is reversible.
In the case of SECR,  there is the stored-energy driving force, 
and 
the reorientation 
proceeds, at least on average, in one direction:
from the orientation of the 'strained crystal' to that of the 'strain-free crystal'.  
Similarly to the shuffles of Merkle, 
some SECR events are much more likely than other, 
and this is manifested by the anisotropy of the interface mobility.

\section{The `nucleation problem'}

The existing theories of recrystallization nucleation explain 
most aspects of the observed phenomena. 
However, there are signs of gaps in the current 
understanding of the early stages of recrystallization.
A long-known issue is related to the 
cube texture component in recrystallization textures,
and another one arises from 
recrystallization experiments on deformed single crystals.

\nextPoint{Inexplicable recrystallization textures}
The assumption of preexistence of embryos at the expected orientations
is at odds with observed recrystallization textures. 
The block theories do not explain how 
orientations absent in the deformed matrix
come to be present in the recrystallized material.
A standard example is that of the 
recrystallization textures of cold-rolled fcc metals. 
They contain the cube texture component.
On the other hand, 
it is known that cube-oriented grains in 
rolled fcc metals are unstable, and there are no such grains in
the deformed state. Therefore, assuming preexistence of cube-oriented embryos, 
one is forced to accept extraordinary rise of some rare grains 
from the background of the deformation texture \cite{Kamijo_1993}. 

It is commonly argued that 
recrystallization starts in high-energy
regions of the crystal, 
in which orientations of the nuclei are 
sufficiently different from that 
of the matrix, e.g. \cite{Cottrell_1953}. In particular, 
it is suggested 
that areas in the needed cube orientation
are present in regions of high orientation gradient, 
e.g., in transition bands \cite{Dillamore_1974}.\footnote{Transition bands, 
i.e., thin zones between regions 
which follow divergent deformation routes,
appear in grains with unstable orientations.
}
Embryos in highly strained regions recover early, and 
those in cube orientation grow because they consume the 
plane-strain texture component $S$  
(as the misorientation between $S$ and the cube orientation is near $M_{g}$;
cf. Fig. \ref{Fig_S_40_111}\textit{a}).

\nextPoint{Experiments on deformed single crystals}
Single crystal and bicrystal experiments of 
Verbraak \cite{Heller_1984} (on Cu bicrystals),
Inoko \cite{Inoko_2004,Inoko_2010} and 
Miszczyk \cite{Miszczyk_2017,Miszczyk_2018,Miszczyk_2018a} (on Al)
clearly indicate that orientations observed in the recrystallized material
are absent from the deformed material. 
Simply put, a single crystal homogeneously deformed in a stable orientation
develops the cell structure with only small deviations of cell orientations 
from that of the crystal, but
upon annealing, new grains in very different orientations appear. 
In the case of Al, 
the new grains are misoriented with respect to the matrix by $M_{g}$ \cite{Miszczyk_2017}.

Clearly, the SIBM mechanism cannot be responsible for the new grains as 
there are no old grains in the new orientations. 
For the same reason, the nucleation via subgrain growth and all low-energy block theories
must be excluded.
It remains to turn to the theories of high-energy blocks.

For claiming  validity of a high-energy block theory, one needs to explain how
light or moderate strains generate large misorientations.
An example of such an explanation is 
Inoko's `rotation recrystallization mechanism';
see \cite{Inoko_2010} and references therein.
According to Inoko, the material in slip-band intersections transforms 
``by the rotation of the deformed matrix around $\langle 111 \rangle$".
It is believed that upon annealing, the region recovers via polygonization, and as
it has high-angle grain boundaries with its surroundings, 
it becomes a viable nucleus.

The problem with the high-energy block theories is that 
modeling of 
plastic deformation predicts relatively large reorientations 
at moderate strains, but they appear 
in shear bands, near grain boundaries and boundary junctions;  
in the case of single crystals deformed in stable orientations,
shear bands and boundaries may be absent.
Moreover, experiments do not show preexisting embryos with orientations observed 
in the recrystallized material. 
In particular, the new orientations are not 
visible on EBSD orientation maps of the deformed single crystals
\cite{Miszczyk_2017,Miszczyk_2018,Miszczyk_2018a}.

\nextPoint{An alternative to the concept of (unobserved) preexisting blocks}
In both cases considered above,
one is forced to assume that
highly strained regions in new orientations 
are beyond resolution of orientation maps. 
This state of affairs was aptly characterized by Humphreys:
"the nucleation stage
is taken to be that part of the recrystallization process which is below 
the resolution limit of the technique being used" \cite{Humphreys_2004}.
The conclusion is that the current theories do not really solve the nucleation problem.

There is, however, another option that is very different from the ones discussed above:
since the new orientations are not observed in the deformed matrix,
one may assume that the  matrix does not contain such orientations.
This alternative to the concept of preexisting high-energy blocks is considered below.

\section{Mechanism of formation of new orientations}

If there are cases in which 
the idea of preexisting embryos in new orientations is 
excluded, 
one must accept that orientations not observed in the matrix
appear ex nihilo via a mechanism which requires an atomic-scale explanation.

\nextPoint{Extended SECR (ESECR)}
As was noted above, strain induced shuffles are likely elements of movement of boundaries 
between the recrystallized material and the deformed matrix. 
If the misorientation of a growing grain is $M_{g}$, the growth is intense, 
and this indicates that the $M_{g}$-related SECR events occur easily. 
The potency of the oriented growth mechanism
means that the SECR mechanism  may be operational 
at early stages of recrystallization. 
The question is how early?

One plausible conjecture is that the orientation-changing shuffles
occur from the very beginning of growth of new crystals. 
It is postulated that SECR, besides participating in the oriented growth,
is capable of transforming 'a strained cluster' to a 'strain-free cluster
in the original or a new orientation'.
By reversing the reasoning, one can argue that 
if the new orientations arise ex nihilo, 
it is likely due to a process not other than that 
responsible for the oriented growth.

\begin{figure}[t]
\begin{picture}(300,250)(0,0)
\put(10,135){\resizebox{7.0 cm}{!}{\includegraphics{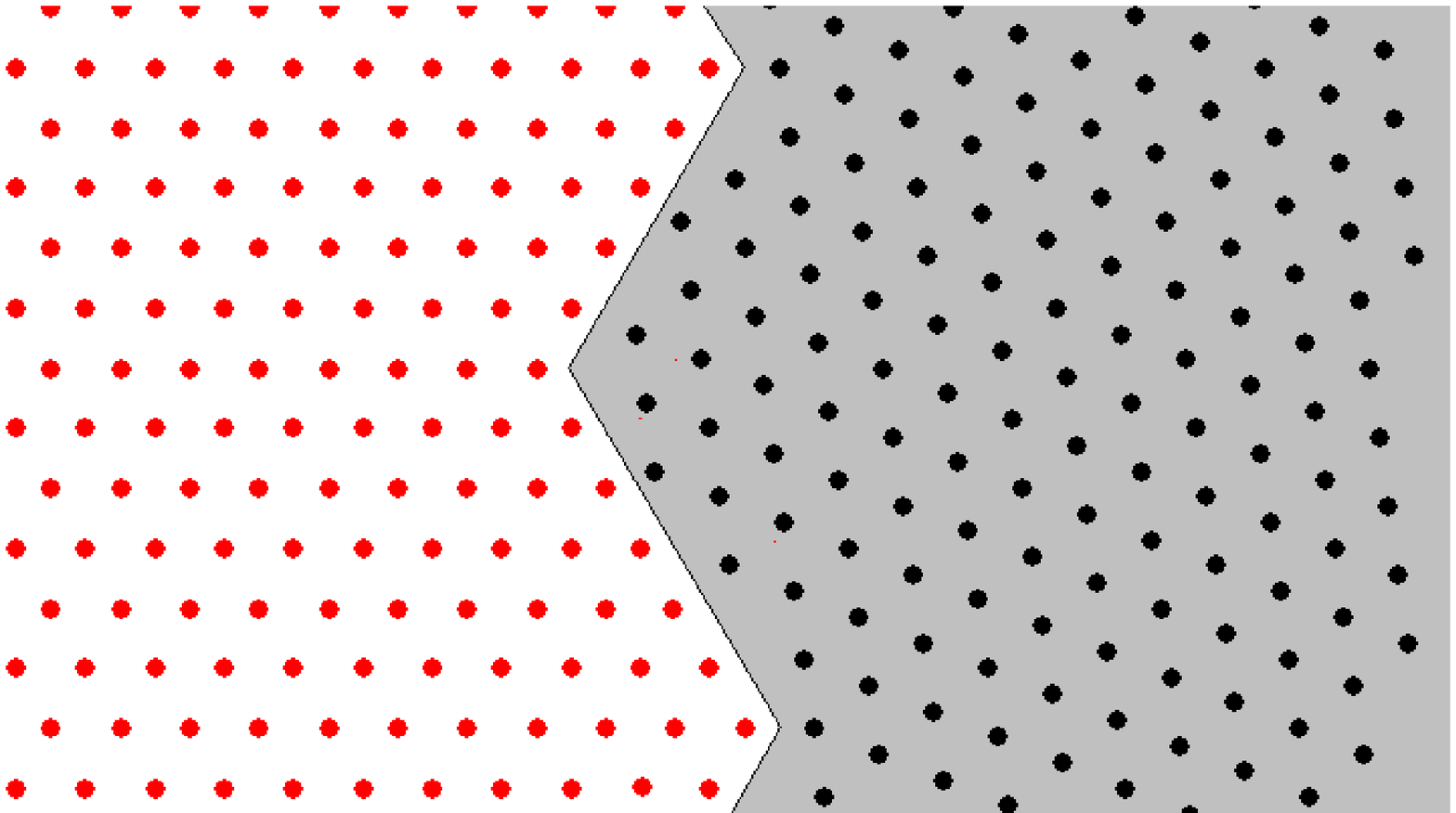}}}
\put(236,135){\resizebox{7.0 cm}{!}{\includegraphics{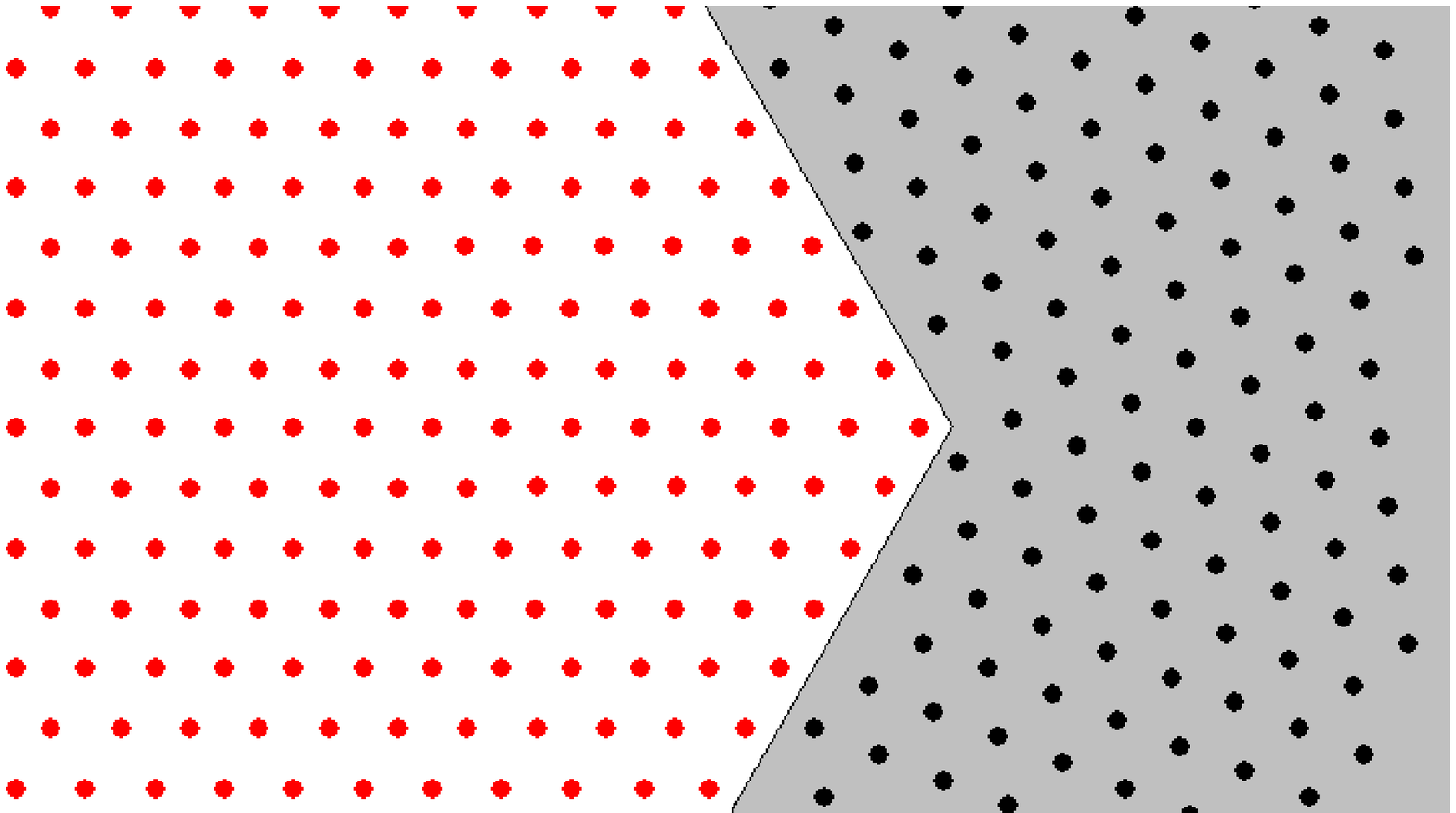}}}
\put(10,0){\resizebox{7.0 cm}{!}{\includegraphics{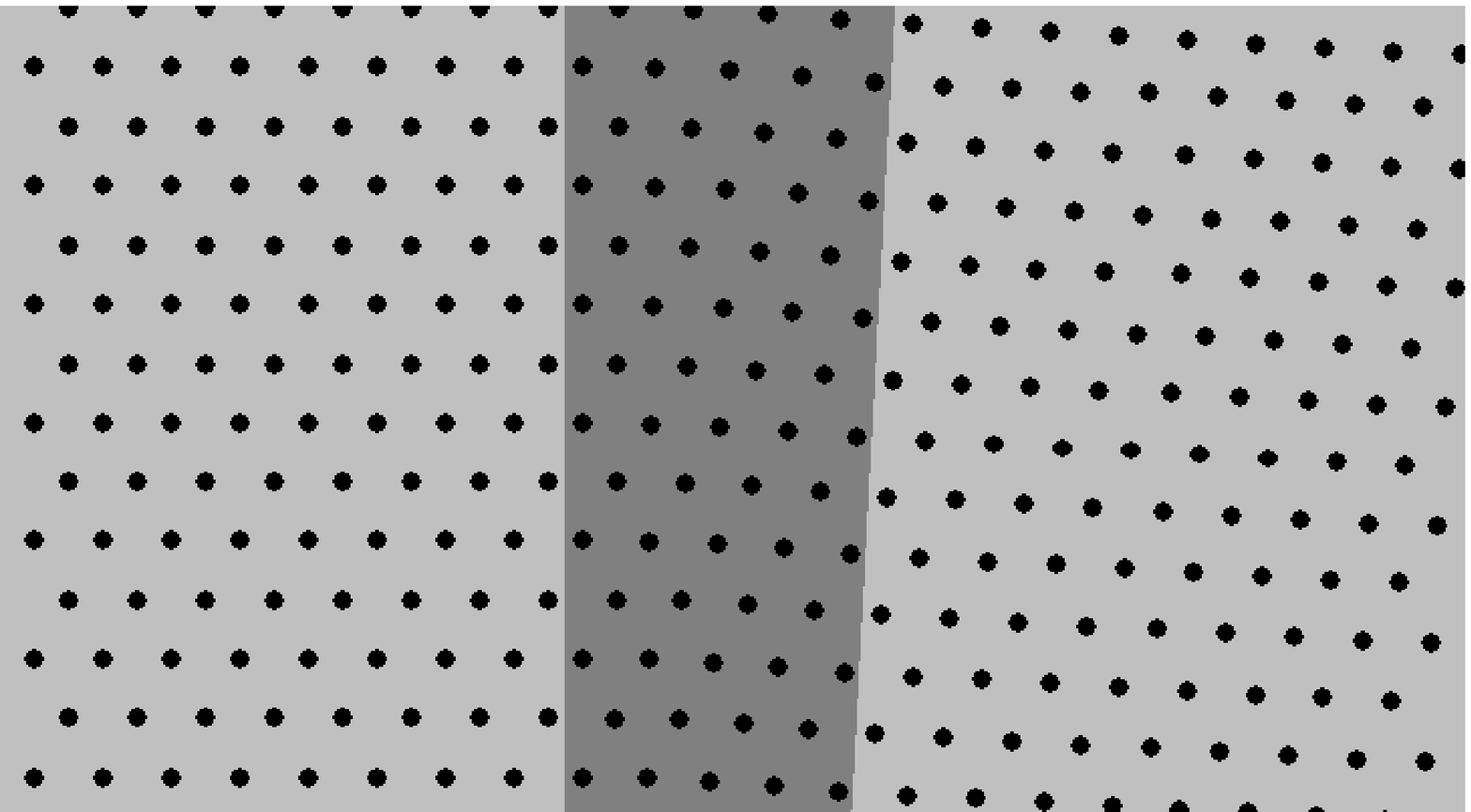}}}
\put(236,0){\resizebox{7.0 cm}{!}{\includegraphics{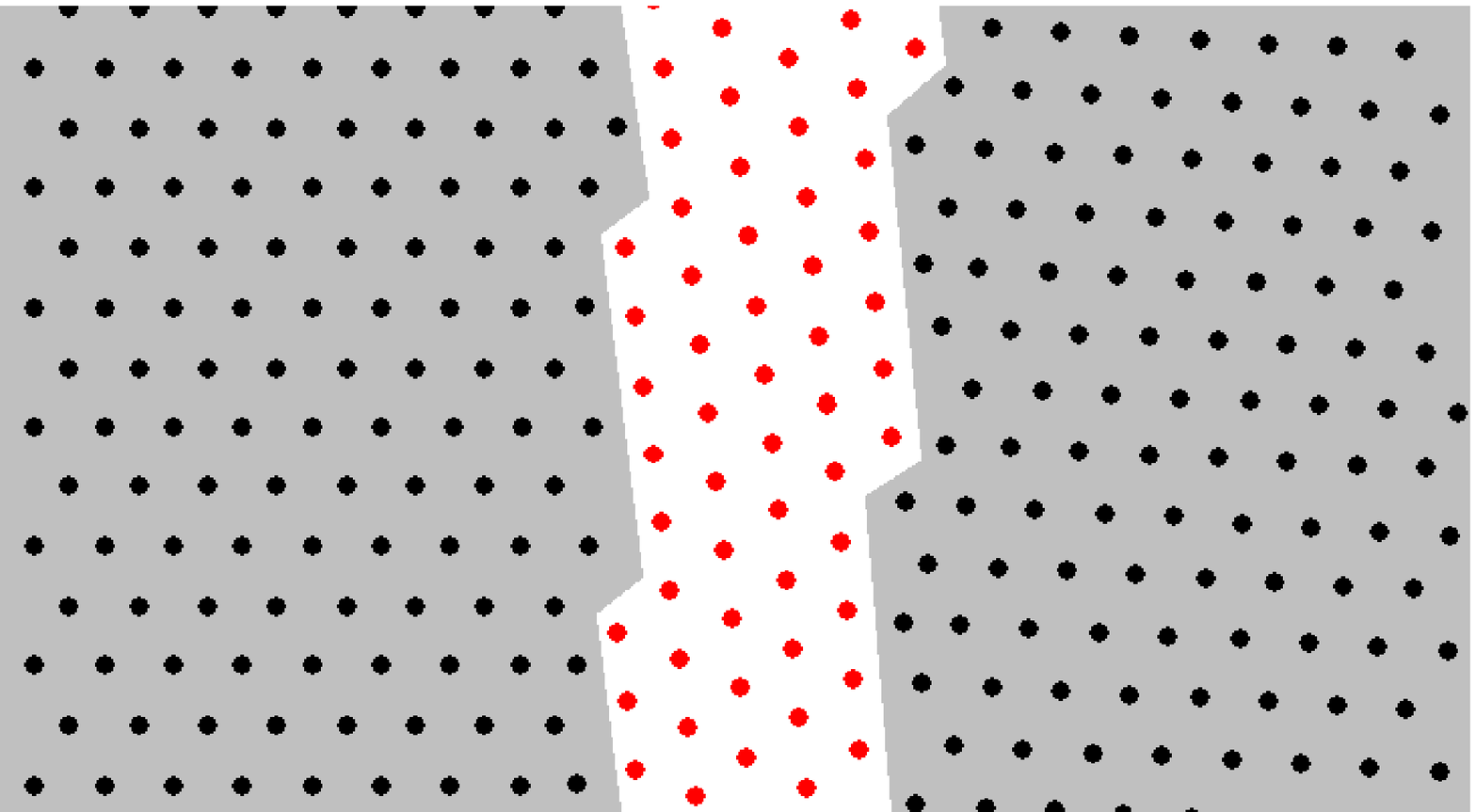}}}
\put(0,240){\textit{a}}
\put(226,240){\textit{b}}
\put(0,103){\textit{c}}
\put(226,103){\textit{d}}
\end{picture}
\vskip 0.0cm
\caption{Schematic illustration of SECR (\textit{a} and \textit{b}) and 
ESECR (\textit{c} and \textit{d}). 
Rearrangement of a cluster of atoms displaces 
the front between the recrystallized matter (white) and 
the deformed matrix (light gray) 
from the position shown in (\textit{a}) to the new one in (\textit{b}).
Atoms in highly strained area marked in dark gray in (\textit{c})
are rearranged, and a new orientation arises (\textit{d}).
}
\label{Fig_SECR}
\end{figure}

The initial shuffling transforming a strained cluster into a strain-free cluster
in the original or a new orientation will be referred to as extended SECR or ESECR 
(Fig. \ref{Fig_SECR}).
It is assumed that 
ESECR may occur in any sufficiently strained region, not necessarily at grain boundaries.

Accounting for ESECR leads to a scenario different from 
that involving nucleation via recovery of preexisting high-energy blocks
and the separate growth stage. 
In the case of shuffle-based mechanism, the stages of nucleation and growth merge: 
the first shuffle leads to a strain-free cluster and 
subsequent shuffles are elements of its growth.

\nextPoint{Nuclation via Rowland transformation}
The idea that new orientations are created by collective atom 
rearrangements has arisen before.
Burgers and Verbraak proposed that nuclei with
new orientations emerge as a result of (inverse) 
Rowland transformation 
causing ``untwinmng'' of interfaces 
between 
regions in twin-related orientations \cite{Burgers_1957}.
The Rowland transformation is a 
collective homogeneous movement of a number of atoms
\cite{Rowland_1955}, 
but unlike the shuffles considered above,
it changes the shape of the transformed cluster
(i.e., it would cause high stresses if the cluster were not very small),
and
the misorientation resulting from the Rowland transformation is distinct 
from the observed $M_{g}$.
Thus, the Burgers-Verbraak mechanism differs from ESECR in that 
the Rowland transformation is not an element of the oriented growth.

\nextPoint{ESECR leading to an old orientation}
The energetically preferred 
ESECR may be the one that leads 
to a strain-free cluster in the orientation of the 
adjacent cell (i.e., in $0$ misorientation).
If the resulting new strain-free cluster turns out to be stagnant,
the shuffle can be seen as an element of the recovery process.
If the new cluster is in a cell wall, 
it can be a precursor to micro-bulging of the cell
in discontinuous subgrain coarsening.
If the new cluster 
in orientation of an adjacent cell is located in an old grain boundary, 
it is a possible precursor to SIBM.
The key element of the `ledge theory', i.e., 
the straightening of a boundary near a ledge, 
can be directly seen as an ESECR event. 
Thus, ESECR leading to an old orientation fits well with the commonly accepted 
mechanisms of early-stage recrystallization.

\nextPoint{ESECR leading to a new orientation}
The ESECR events leading to new orientations 
can be viewed as faults in the recovery process.
With the formation of new orientations and selective growth 
governed by the same mechanism, 
the orientation selectivity in
creation of new grains is likely to be related to 
the selectivity of growth;
the more selective the oriented growth, 
the more restricted the orientations of the emerging new crystallites.
With highly selective growth,
ESECR events causing misorientations other than the $0$ misorientation and $M_{g}$ are 
expected to be infrequent and such clusters have little chance to grow. 
With the postulated mechanism, the `oriented nucleation'  meets the `oriented growth'.

In low stacking fault energy fcc metals, 
a shuffle of atoms can be a part of annealing twinning.
In the case of Al, the $\Sigma 3$ misorientations are rarely observed, which means 
that $\Sigma 3$-related ESECR and SECR events
are energetically less favorable than those linked to $0$ and $M_{g}$.

\nextPoint{ESECR and conventional nucleation}
ESECR can also be seen as the first step in 
formation of a nucleus.
Thus, the ESECR clusters in new orientations are subject to the rules 
of the conventional nucleation theory; 
capillary forces are likely to switch  
a subcritical cluster back to the original orientation.
Thus, the arguments used for rejecting the conventional nucleation mechanism 
are applicable. 
However, there are reasons to doubt these arguments.

The strength of the oriented growth indicates that 
when the misorientation is $M_{g}$, 
the ratio of areal surface energy and the volumetric (difference in) stored energy
is low, 
which means that the critical radius of $M_{g}$-misoriented nuclei is small. 
This opens up the possibility of the formation of such nuclei.
Already Bailey \cite{Bailey_1960} admitted 
that local dislocation density (e.g., in cell walls) may be high enough 
for the critical radius to be reasonably small, but
he noted that 
the growth of a nucleus 
would be arrested once it grows beyond the wall area into the cells
where the density is low. But is such arrest truly inevitable? 
The nucleus does not have to be spherical; a plate-like structure can grow within the wall;
see Appendix.
The point is that the formation of new grains by conventional nucleation 
is not completely ruled out if the theory is extended 
to account for the intricacies of recrystallization.

The above argument falls within the scope of conventional nucleation theory.
However, there are also reasons to question 
the applicability of this theory to recrystallization
as there are significant differences between 
conventional nucleation 
from a supersaturated solution 
and early stages of recrystallization. 
The conventional nucleation 
(initially developed for description of condensation of vapors into liquids)
has purely spontaneous or stochastic character,
whereas new grains in recrystallization 
are not random in terms of their locations
and far from random in terms of their orientations. 
Clearly, the conventional theory gives no information on the 
orientation of the new crystallites
or on paths from old to new orientations.

In the case of crystallites, 
the surface tension is anisotropic and depends on surface curvature. 
Anisotropy causes faceting, and
particular facets may have different energies and different kinetic properties.
Small crystallites are beyond the simple mesoscopic descriptions
based on the assumption of isotropy.

In conventional nucleation,
a subcritical cluster is ephemeral: it may either grow or decay to the initial 
state.
In recrystallization, 
a sufficiently large but still subcritical cluster formed in the deformed region 
may also grow or decay, but the decay may lead not to the initial deformed state, 
but to a less strained (possibly strain-free) 
domain. 
The conventional theory of nucleation does not cover such scenarios.

\section{Concluding remarks}

There are a number of hard facts about recrystallization.
Above all, the phenomenon of oriented growth has been 
repeatedly supported by experimental observations.
Certainly, in polycrystalline samples, 
new grains may appear via SIBM or discontinuous subgrain coarsening. 
Moreover, there are circumstances which cannot be explained by 
standard recrystallization nucleation theories recognized today.
These theories assume that orientations of recrystallized grains 
preexist in the deformed material,
while in some cases there are strong indications for the absence of these orientations
in the deformed matrix.  

To address this issue, it has been proposed earlier that 
new orientations arise by recovery of preexisting high-energy blocks.
Since such blocks are not observed on EBSD orientation maps, 
their size must be smaller than the resolution of the mapping method.

An alternative explanation proposed in this paper is that the 
new orientations arise ex nihilo via 
a mechanism akin to that involved in boundary migration leading to oriented growth. 
The strength of the oriented growth mechanism indicates that it operates at 
an early stage of grain evolution. 
Since mobility of boundaries under the stored-energy driving force is anisotropic,
the boundaries most likely migrate by shuffles, i.e., 
collective rearrangements of atoms.
It is postulated that shuffles not only contribute to the oriented growth,
but can also transform a strained cluster of atoms to a strain-free domain
in a different orientation -- a nucleus of a growing grain.

The proposed mechanism of formation of new orientations
is only a hypothesis which needs to be verified. 
However, the means to test it are limited.
One may look for missing orientations in the deformed matrix 
using high-resolution orientation mapping methods 
or the dark field electron microscopy technique; 
the problem is that usually the higher the resolution, 
the smaller the observed area.
The hypothesis could be partly supported or falsified 
by molecular dynamics simulations.

The presented description of the proposed mechanism
leaves many questions unanswered.
For instance, assuming validity of the mechanism,
at which conditions would it 
be active?
Why the particular misorientation of new grains stands out?
What is the character of strain leading to shuffles; 
is it large dislocation content or large elastic strain? 
What are the actual displacements of atoms in individual shuffles?
What are the sizes of clusters affected by the shuffles?
How do the shuffles break down over time?
If the hypothesis is confirmed, these questions will need to be addressed.

There are a number of reasons for seriously considering the described mechanism. 
It explains the formation of new orientations.  
It 
establishes the beginning of the oriented growth. 
The important aspect of the mechanism is its simplicity;
the collective shuffle of atoms is the single key constituent of both 
the raise of new grains and their subsequent growth.
In a sense, the mechanism merges oriented nucleation and oriented growth.
The presented hypothesis has the potential to 
shift the paradigm of recrystallization nucleation
and change the narrative 
about the early stages of recrystallization.



\newpage

\newpage
\bibliographystyle{unsrt}
\bibliography{Rec_nucl} 

\newpage

\section{Appendix}

\nextPoint{Variant selection} 
Due to crystal symmetry, 
a misorientation 
corresponds 
to a number of symmetrically equivalent rotations. 
For a given orientation of the deformed matrix, these rotations 
lead to distinct orientations of the recrystallized grains
known as variants. 
The frequency of occurrence of the variants may be 
affected by external factors (linked to the specimen reference frame);
some variants can be less frequent than others. 
The variant selection for recrystallization growth 
was observed by Kohara et al. \cite{Kohara_1958b}.
The occurrence of variant selection 
indicates that the boundary mobility depends not only on 
the macroscopic boundary parameters (misorientation and boundary plane)
but also on boundary orientation with respect to the external reference frame.

\begin{figure}[t]
\begin{picture}(300,290)(0,0)
\put(40,0){\resizebox{13.0 cm}{!}{\includegraphics{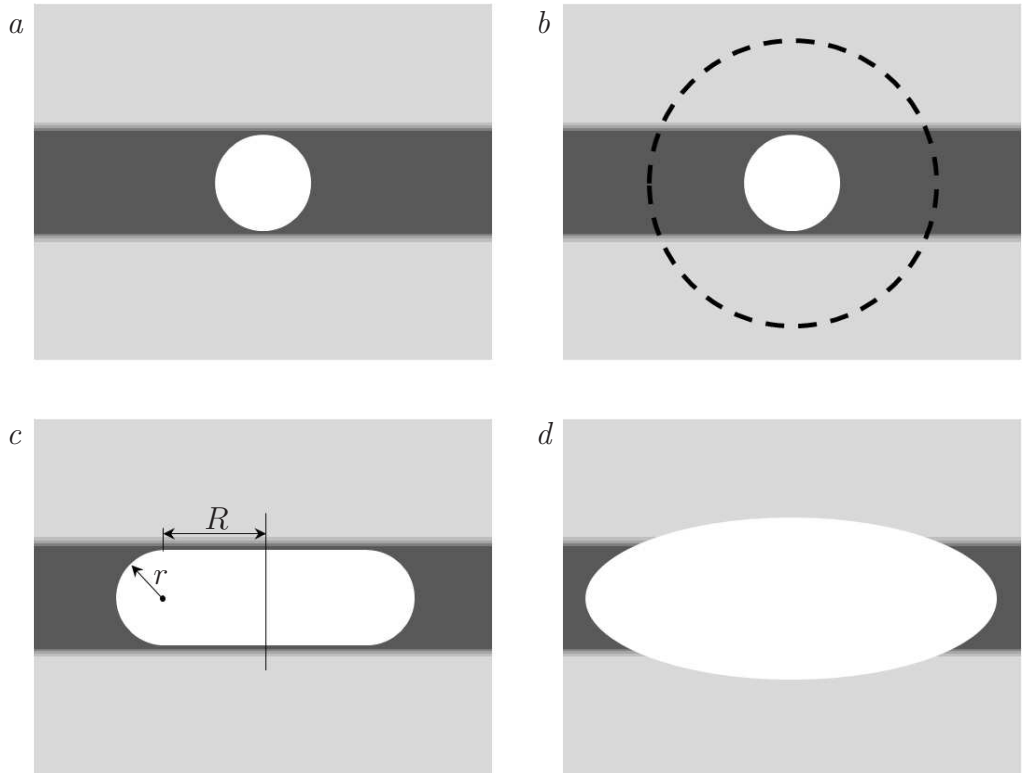}}}
\put(30,280){\textit{a}}
\put(228,280){\textit{b}}
\put(30,125){\textit{c}}
\put(228,125){\textit{d}}
\put(85,72){$r$}
\put(104,93){$R$}
\end{picture}
\vskip 0.0cm
\caption{Schematic of the model of nucleation in cell walls
with high dislocation density.
(\textit{a}) The defect free nucleus (white) in a cell wall (dark gray) 
and interiors of the cells (light gray). 
(\textit{b}) When growing, the nucleus cannot keep its spherical shape
because the stored energy density inside the cells is too low.
(\textit{c}) The criterion for growth is satisfied by the oblate solid 
of revolution if $r > r_c$.
(\textit{d}) The low curvature interfaces of the solid
can move into the cells due to a (small) difference in stored energy. 
}
\label{Fig_pancake}
\end{figure}

\nextPoint{Aspherical nucleus in a cell wall}
The description of a nucleus in a cell wall is complicated as it involves 
three regions with different densities of stored energies and three  
interfaces with different energies. 
One can deal with a simplified model with 
the high dislocation density wall separating two low dislocation density cells;
see Fig. \ref{Fig_pancake}\textit{a}.
Let $\gamma_S$ and $g_V$ denote densities of 
the free energy of the wall-nucleus interface and 
the difference of energies stored in the wall and the nucleus, respectively.
The wall thickness is assumed to be of the order of the 
critical diameter $2 r_c = 4 \gamma_S/g_V$, 
and energies of all interfaces are 
assumed to be of the same magnitude $\gamma_S$.  
As was noted by Bailey \cite{Bailey_1960}, a spherically shaped nucleus could not grow 
(Fig. \ref{Fig_pancake}\textit{b}).
However, what may happen is the aspherical growth within the wall. 
Take the solid of revolution obtained by rotating the figure built of 
two semi-disks attached to opposite sides of a rectangle
about the symmetry axis perpendicular to the straight sides (Fig. \ref{Fig_pancake}\textit{c}). 
The volume of the solid and its surface area are, respectively, 
$V_n = 4 \pi r^3/3 + \pi R r \left(2 R+ \pi  r \right)$
and
$S_n=4  \pi r^2 + 2 \pi  R (R+\pi  r)$,
where $r$ is the radius of the semi-disks and $2R$ is the distance between
their centers. 
Based on the above expressions, one can show that 
if $r > r_c$, the increase of $R$ entails a drop of the total energy $\gamma_S S_n - g_V V_n$,
i.e., the nucleus with fixed $r$ can grow within the wall by increasing $R$.
In consequence, it will develop interfaces of curvature 
so low that they can grow into the cells despite 
the smallness of the difference in stored energy
(Fig. \ref{Fig_pancake}\textit{d}). 
Clearly, the actual picture is much more complicated.
The point is that even in such a simplified case, 
the thermally-activated nucleation can be envisioned.
Additional complications (e.g., wall junctions) 
make such nucleation more likely.

\end{document}